# A periodic level-crossing two-state model of a general Heun class

G. Saget[1], A.M. Ishkhanyan[2,3], C. Leroy[1] and T.A. Ishkhanyan[1,4]

[1]Laboratoire Interdisciplinaire Carnot de Bourgogne, UMR CNRS 6303-Université de Bourgogne Franche-Comté, Dijon, 21078 France
[2]Institute for Physical Research, NAS of Armenia, Ashtarak, 0203 Armenia
[3]Armenian State Pedagogical University, Yerevan, 0010 Armenia
[4]Moscow Institute of Physics and Technology, Dolgoprudny, 141700 Russia

**Abstract.** We present a specific constant-amplitude periodic level-crossing model of the semi-classical quantum time-dependent two-state problem that belongs to a general Heun class of field configurations. The exact analytic solution for the probability amplitude, generally written for this class in terms of the general Heun functions, in this specific case admits series expansion in terms of the incomplete Beta functions. Terminating this series results in an infinite hierarchy of finite-sum closed-form solutions each standing for a particular two-state model, which generally is only *conditionally* integrable in the sense that for these field configurations the amplitude and phase modulation functions are not varied independently. However, there exists at least one exception when the model is *unconditionally* integrable, that is the Rabi frequency and the detuning of the driving optical field are controlled independently. This is a constant-amplitude periodic level-crossing model, for which the detuning in a limit becomes a Dirac delta-comb configuration with variable frequency of the level-crossings. We derive the exact solution for this model, determine the Floquet exponents and study the population dynamics in the system for various regions of the input parameters.

## 1. Introduction

The two-state system is the simplest non-trivial quantum system that can be treated analytically and it is a rather good approximation for many real quantum systems in nature. Examples of such systems arise in a number of phenomena of contemporary physics, chemistry, engineering, etc. In the present paper we consider the quantum non-adiabatic transitions between levels [1] during the interaction of external optical fields with the matter, when only a transition between two of the levels of a generally multi-state system is nearly resonant with the external driving field. So, we discuss the dynamics of just two quantum levels, i.e., the quantum two-state problem [1-2], assuming the effect of the field on other levels to be negligible as they are far off the resonance.

The specific model we consider presents a constant-amplitude periodic phase-modulation field configuration, which may be level-crossing, level glancing or non-crossing.

We note that though the level-crossing is a key paradigm of the theory of quantum non-adiabatic transitions [1-2], only a few analytic models describing such processes are known [3-9]. Besides, these are only single-crossing models, and to the best of our knowledge no exactly solvable models of periodic crossings of the resonance are known.

The Schrödinger equation for a quantum two-state system in semi-classical time-dependent formulation is written as a system of two coupled first-order ordinary linear differential equations for the probability amplitudes of the two involved states. This system is further reduced to one second-order differential equation for the probability amplitude of the lower or upper level.

A basic set of analytic models of the two-state problem has been developed in the past by solving the time-dependent Schrödinger equations in terms of special functions of the hypergeometric class [3-14]. More models are derived when expressing the solution of the problem in terms of the Heun functions that are solutions of the five Heun equations [15-17]. It should be noted that the Heun functions, which compose the next generation of special functions, represent direct generalizations of the hypergeometric functions. For this reason, the solutions in terms of the Heun functions generalize all known hypergeometric cases.

We have recently shown that there exist in total 61 infinite classes of two-state models solvable in terms of the five Heun functions [18-20]. There are thirty-five classes for which the problem is exactly solved in terms of the general Heun functions [18], fifteen classes solvable in terms of the single-confluent Heun functions [19], five classes in terms of the double-confluent Heun functions [20], five other classes in terms of the bi-confluent Heun functions [20,21], and a class in terms of the tri-confluent Heun functions [20].

In the present paper we discuss a constant-amplitude periodic level-crossing model belonging to one of the thirty-five general Heun classes of field configurations [18]. In the particular case at hand the general Heun function involved in the solution of the two-state problem admits a series expansion in terms of the incomplete Beta functions. The coefficients of the expansion obey a three-term recurrence relation which allows termination of the series. We show that the conditions for simultaneous left- and right-hand side terminations generally lead to *conditionally* integrable models for which the amplitude- and detuning-modulation functions are not varied independently. However, there exists a particular *unconditionally* integrable model. This is a constant-amplitude periodic level-crossing model for which the detuning modulation function for a large parameter is effectively a Dirac delta-comb threaded on a carrier frequency. Notably, the exact solution of the problem for this particular model is eventually written in terms of elementary functions. This solution explicitly indicates the Floquet exponents expressed through the generalized Rabi frequency for the carrier frequency of the associated constant-detuning field. Using the derived solution, we explore the non-adiabatic dynamics of the two-state system subject to excitation by a driving optical field of the mentioned configuration.

## 2. A constant-amplitude level-crossing general Heun model of the two-state problem

The governing equations for the probability amplitudes $a_1(t)$ and $a_2(t)$ of the ground and excited states, respectively, for the semi-classical time-dependent two-state problem in the rotating-wave approximation are written as the following system of the first-order linear ordinary differential equations [1,2]:

$$i\frac{da_1(t)}{dt} = U(t)e^{-i\delta(t)}a_2(t), \quad i\frac{da_2(t)}{dt} = U(t)e^{+i\delta(t)}a_1(t). \tag{1}$$

Here the amplitude modulation function $U(t)$ (the Rabi frequency) and the frequency modulation function $\delta(t)$ (the derivative $\delta_t = d\delta/dt$ of this function is the detuning of the transition frequency from the field frequency) are arbitrary real functions of time. The system is readily reduced to one second-order linear differential equation written for one of the probability amplitudes. For the excited state's probability amplitude $a_2(t)$ the equation reads

$$a_{2tt}(t) + \left(-i\delta_t(t) - \frac{U_t(t)}{U(t)}\right)a_{2t}(t) + U(t)^2 a_2(t) = 0, \tag{2}$$

where (and hereafter) the lowercase Latin alphabetical index denotes differentiation with respect to the corresponding variable.

The reduction of equation (2) to the general Heun equation [15-17]

$$u_{zz} + \left(\frac{\gamma}{z} + \frac{\delta}{z-1} + \frac{\varepsilon}{z-a}\right)u_z + \frac{\alpha\beta z - q}{z(z-1)(z-a)}u = 0 \tag{3}$$

leads to thirty-five classes of field configurations given as [18]

$$U(t) = U_0^* z^{k_1}(z-1)^{k_2}(z-a)^{k_3}\frac{dz}{dt}, \tag{4}$$

$$\delta_t(t) = \left(\frac{\delta_1}{z} + \frac{\delta_2}{z-1} + \frac{\delta_3}{z-a}\right)\frac{dz}{dt}, \tag{5}$$

where $z = z(t)$ is an arbitrary complex-valued function of time and $k_{1,2,3}$ are integers or half-integers obeying the inequalities $-1 \leq k_{1,2,3} \cup k_1 + k_2 + k_3 \leq -1$. The parameters $U_0^*$ and $\delta_{1,2,3}$ are generally complex constants that should be chosen so that the functions $U(t)$ and $\delta(t)$ are real for the chosen transformation $z(t)$. Since the involved parameters including the third singular point $a$ of the general Heun equation (3) are arbitrary, all thirty-five classes are in general at least five-parametric (more parameters may come from the transformation $z(t)$).

The two-state model that we introduce belongs to the general Heun class with $k_{1,2,3} = (-1,0,0)$. The model is given through the equations (4),(5) with the complex-valued transformation of the independent variable taken as $z(t) = \sqrt{a}\, e^{i\Delta(t-t_0)}$ [18]. With the choice of the parameters of the amplitude and detuning modulation functions as

$$U_0^* = -i\frac{U_0}{\Delta}, \quad \delta_1 = -i\frac{\Delta_1}{\Delta} \quad \delta_2 = -\delta_3 = i\frac{\Delta_2}{\Delta}, \tag{6}$$

we get a 6-parametric constant-amplitude field configuration with periodic modulation of the detuning given as (see equations (46) and (47) in [18])

$$U(t) = U_0, \quad \delta_t(t) = \Delta_1 + \frac{(1-a)\Delta_2}{1+a-2\sqrt{a}\cos(\Delta(t-t_0))}, \tag{7}$$

where $U_0, a, \Delta_1, \Delta_2$ and $\Delta, t_0$ are arbitrary real input parameters, $a > 0$. Different particular values of these input parameters produce both level-crossing and non-crossing detuning modulations, as well as level-glancing configurations (figure 1). Since the detuning is a $T = 2\pi/\Delta$-periodic function of time, the level-glancing occurs if the condition for resonance $\delta_t(t) = 0$ is achieved at an extremum of the function $\cos(\Delta t)$, that is if $t = t_0$ or $t = t_0 + \pi/\Delta$.

This is the case if
$$\frac{\Delta_1}{\Delta_2} = \frac{\sqrt{a}+1}{\sqrt{a}-1} \quad \text{or} \quad \frac{\Delta_1}{\Delta_2} = \frac{\sqrt{a}-1}{\sqrt{a}+1}. \tag{8}$$

The level-glancing sub-family of detuning achieved by the second choice is shown in fig. 2.

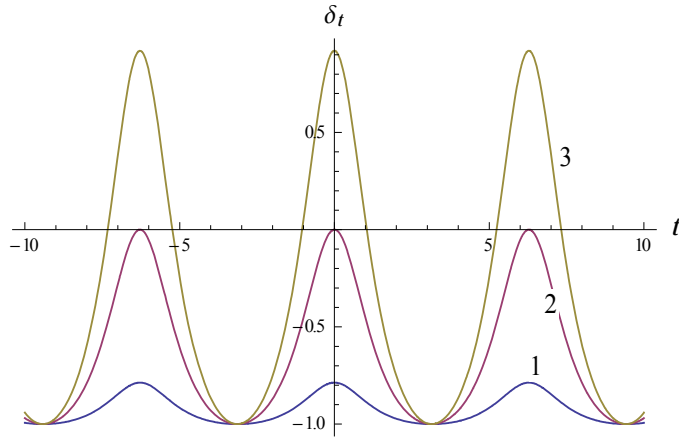

Fig.1. Detuning modulation function (7) for $a = 16$ and $\Delta_1 = -1 + 3\Delta_2/5$. $\Delta_2 = -0.2, -15/16, -1.8$ for curves 1,2,3, respectively, $\Delta = 1$, $t_0 = 0$.

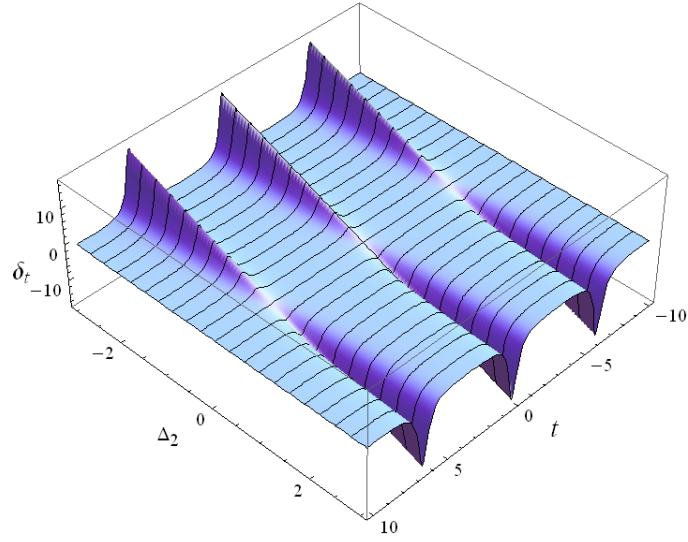

Fig.2. A level-glancing sub-family of the detuning modulation function (7):
$\Delta_1 = (\sqrt{a}-1)/(\sqrt{a}+1)\Delta_2$, $a = 2$ ($\Delta = 1$, $t_0 = 0$).

The solution of the two-state problem (2) is explicitly written as

$$a_2 = C_0 z^{\alpha_1}(z-1)^{\alpha_2}(z-a)^{\alpha_3} H_G(a,q;\alpha,\beta;\gamma,\delta,\varepsilon;z), \qquad (9)$$

where $C_0$ is an arbitrary constant, the parameters $\gamma, \delta, \varepsilon, \alpha, \beta, q$ of the general Heun function $H_G$ are determined through the equations (15)-(17) of [18] as (without loss of generality we put $\Delta = 1$)

$$(\gamma, \delta, \varepsilon, \alpha, \beta, q) = \left(1 \pm \sqrt{4U_0^2 + \Delta_1^2}, \Delta_2, -\Delta_2, 0, \pm\sqrt{4U_0^2 + \Delta_1^2}, (a-1)\Delta_2\alpha_1\right), \qquad (10)$$

and the pre-factor parameters $\alpha_{1,2,3}$ are given as

$$\alpha_1 = \frac{\Delta_1}{2} \pm \sqrt{U_0^2 + \frac{\Delta_1^2}{4}}, \quad \alpha_{2,3} = 0. \qquad (11)$$

We note that the plus and minus signs in the expression of $\alpha_1$ produce two independent fundamental solutions.

The general Heun function is a rather complicated mathematical object the theory of which is currently poorly developed. However, during the past years a progress is recorded following the approach suggested by Svartholm [22] and Erdélyi [23]. Several new series expansions of the general Heun function have been constructed in terms of simpler special functions such as the incomplete Beta function, the Gauss hypergeometric function, the Appell generalized hypergeometric function of two variables [24-26]. Below we use a specific expansion of the general Heun function which is applicable if a characteristic exponent of the singularity at infinity is zero [24]. As it is seen from equation (10), in our case the last condition is satisfied as $\alpha = 0$.

## 3. Series solutions of the Heun equation in terms of the incomplete Beta functions

Thus, for the field configuration (7) a characteristic exponent of the regular singularity of the general Heun equation (3) at infinity is zero. It has been shown in [24] that the Heun function then permits a series expansion in terms of the incomplete Beta functions:

$$u = \sum_n c_n u_n, \quad u_n = B_z(\gamma_n, \delta_n). \tag{12}$$

This expansion is developed as follows. The involved Beta functions satisfy the following second-order linear differential equation:

$$\frac{d^2 u_n}{dz^2} + \left(\frac{1-\gamma_n}{z} + \frac{1-\delta_n}{z-1}\right)\frac{du_n}{dz} = 0. \tag{13}$$

We note that this is a particular specialization of the Gauss hypergeometric equation for which at least one of the characteristic exponents is zero as it is the case for the Heun equation for the field configuration (7). If we now put $\delta_n = 1 - \delta$ for all $n$, we will then make the characteristic exponents coincide at the singular point $z = 1$ as well. Then, the strategy is to achieve the correct behavior of function (12) at the remaining singularities $z = 0$ and $z = a$ by adjusting the parameters $\gamma_n$ and the coefficients $c_n$ of the sum. This is done by substitution of equations (12) and (13) into equation (3), and then, using the recurrence relations between the involved Beta-functions, grouping the terms proportional to a particular $u_n$ and finally requiring all the resultant summands of thereby regrouped sum to vanish. After some algebra we arrive at a three-term recurrence relation for successive coefficients:

$$R_n c_n + Q_{n-1} c_{n-1} + P_{n-2} c_{n-2} = 0, \tag{14}$$

where

$$R_n = a(\gamma - 1 + \gamma_n)(\gamma_n - 1), \tag{15}$$

$$Q_n = -a(\gamma - 2 + \gamma_n)(\gamma_n + \delta_n - 2) - ((\gamma - 2 + \gamma_n) + \varepsilon(\gamma_n - 1)) - q, \tag{16}$$

$$P_n = (\gamma_n + \delta_n - 2)((\gamma - 3 + \gamma_n) + \varepsilon). \tag{17}$$

Assuming $\gamma_{n\pm 1} = \gamma_n \pm 1$, for the left-hand side termination of the series at $n = 0$ it should hold $R_0 = 0$, so that $\gamma_0 = 1 - \gamma$ or $\gamma_0 = 1$. The choice $\gamma_0 = 1$ does not work since then $u_{-1} = B(z, 0, \delta_n)$ which is not defined, so the condition for the left-hand side termination is

$$\gamma_0 = 1 - \gamma. \tag{18}$$

We note that this means that the characteristic exponents of equation (12) at the singular point $z = 0$ for the first term of the expansion with $n = 0$, that is for the term $u_0 = B_z(\gamma_0, \delta)$, are also the same as those for the Heun equation (3).

Thus, we finally have the expansion

$$u = \sum_{n=0}^{\infty} c_n B_z(1 - \gamma + n, 1 - \delta) \tag{19}$$

with the coefficients of the recurrence relation (14) being simplified as

$$R_n = an(n - \gamma) \tag{20}$$

$$Q_n = -an(n + 1 - \gamma - \delta) - (n + \varepsilon)(n + 1 - \gamma) - q, \tag{21}$$

$$P_n = (n + 2 - \gamma - \delta)(n + \varepsilon). \tag{22}$$

The series (19) terminates from the right-hand side and thus generates a closed-form finite-sum solution if two successive coefficients, say $c_{N+1}$ and $c_{N+2}$, vanish for some $N = 0, 1, 2, \ldots$. Equation $c_{N+2} = 0$ is satisfied if $P_N = 0$. This is equivalent to the condition

$$\varepsilon = -N, \ N = 0, 1, 2, \ldots \tag{23}$$

or
$$\gamma + \delta - 2 = +N, \ N = 0, 1, 2, \ldots, \tag{24}$$

while the equation $c_{N+1} = 0$ (this equation is referred to as the $q$-equation) leads to a polynomial equation of the degree $N + 1$ for the accessory parameter $q$ thus imposing a restriction on this parameter.

## 4. Conditionally and unconditionally integrable sub-models

Having expanded the general Heun function into series in terms of the incomplete Beta-functions and further established the conditions for termination of the constructed series, we now consider if the solution of the two-state problem for the field configuration (7) can be written through a linear combination of a finite number of incomplete Beta functions. So we inspect if the termination conditions are satisfied for the parameters of the general Heun function given by equation (10). It is then found out that for real input parameters $U_0, a, \Delta_1, \Delta_2$ the termination is not achieved if equation (24) is applied. For the alternative condition $\varepsilon = -N$ given by equation (23), however, the answer is positive for all non-negative integers $N = 0, 1, 2, 3 \ldots$. For $N = 0$ and $N = 1$ we get the trivial constant-detuning Rabi model. However, starting from $N = 2$, the results become non-trivial. It is then understood that we thereby derive an infinite hierarchy of particular sub-models for which the solution of the two-state problem is given by a linear combination of a finite number of incomplete Beta functions. Consider these cases in more detail.

To meet the termination condition $\varepsilon = -N$ suggested by equation (23), we put $\Delta_2 = N$ (see equation (10)) with a non-negative integer $N = 0,1,2,...$. This reduces the number of variable input parameters of the field configuration (7) to three (we omit the time scaling and shifting parameters $\Delta$ and $t_0$). The second termination condition $c_{N+1} = 0$ will of course further decrease the number of the independent parameters by imposing a relation between the remaining parameters $U_0, \Delta_1$ and $a$. If this is a relation involving only the detuning parameters $\Delta_1$ and $a$, then we have an *unconditionally* integrable model for which the frequency detuning and the Rabi frequency are independent. Otherwise, if the relation links the detuning parameters with the Rabi frequency $U_0$, the model is called *conditionally* integrable. The result is that for $N = 2$ we have an unconditionally integrable model (this model is presented in the next section), while for $N = 3$ the model is proved to be conditionally integrable. It is expected that for all higher orders $N > 3$ the models are also conditionally integrable. To give an explicit example of such models, here is the field configuration for the case $N = 3$:

$$U = U_0, \quad \delta_t = \Delta_1 + \frac{9 - 3\sqrt{3}R - 9\Delta_1}{\left(\sqrt{3} - R\right)R + 3(\Delta_1 - 1)\Delta_1 + \sqrt{1 - \frac{6}{3 + \sqrt{3}R - 3\Delta_1}}\left(R^2 - 3(\Delta_1 - 1)^2\right)\cos(t)}, \quad (25)$$

where
$$R = \pm\sqrt{U_0^2 + \Delta_1^2 - 1}. \quad (26)$$

## 5. The unconditionally integrable sub-model

Consider the case $N = 2$, i.e. when the general Heun function involved in solution (9) is represented as a sum of three incomplete Beta functions. We have $\Delta_2 = 2$ and the equation $c_{N+1} = 0$ in terms of the input physical parameters of the problem $U_0, \Delta_1$ and $a$ (we assume $\Delta = 1$) is reduced to

$$U_0^2(a-1)^2\left(a(\Delta_1 - 1) - \Delta_1 - 1\right) = 0. \quad (27)$$

Since $U_0 \neq 0$ and $a \neq 1$, this equation relates the parameters $a$ and $\Delta_1$:

$$a = \frac{\Delta_1 + 1}{\Delta_1 - 1}, \quad (28)$$

thus generating an *unconditionally* solvable 2-parametric two-state model:

$$U(t) = U_0 = \text{const}, \quad \delta_t(t) = \Delta_1 - \frac{2}{\Delta_1 - \sqrt{\Delta_1^2 - 1}\cos(t)}. \tag{29}$$

We note that since the detuning should be real, the model is applicable for $|\Delta_1| > 1$.

The 3D-plot of the detuning modulation function (29) depending on the parameter $\Delta_1$ is shown in figure 3. The plot of this function for several particular values of $\Delta_1$ depicted in figure 4 shows that with increasing $\Delta_1$ the extremum of the detuning approaches $\Delta_1$, so that for large $\Delta_1$ the detuning modulation function effectively becomes a periodic Dirac delta-comb threaded on the line $\delta_t(t) = \Delta_1$.

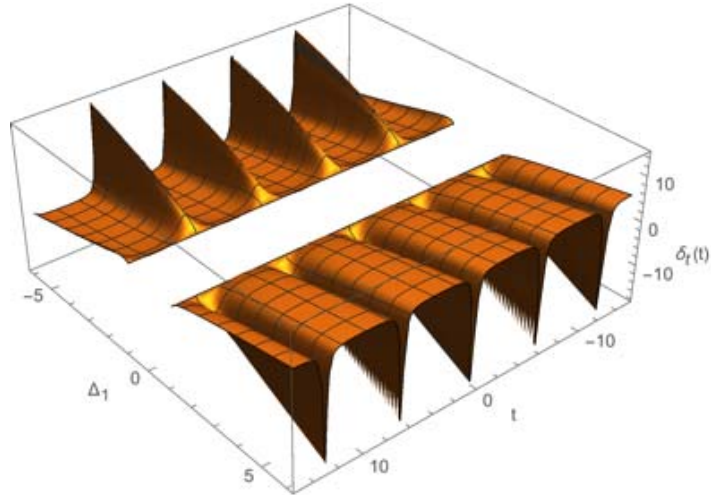

Fig. 3. Exactly integrable constant-amplitude periodic-crossing model (29): 3D plot of the detuning $\delta_t(t)$.

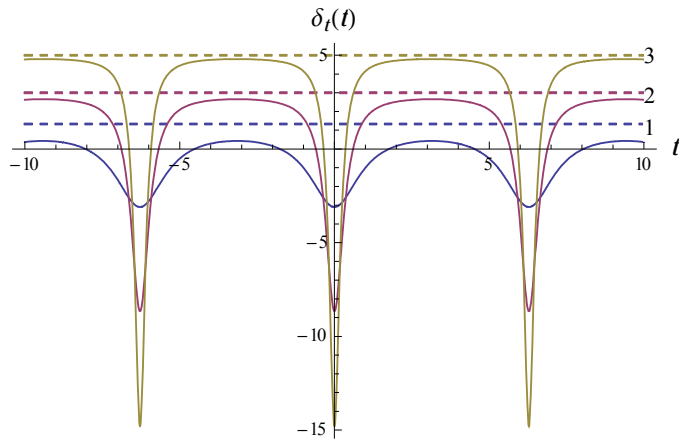

Fig. 4. Exactly integrable constant-amplitude periodic-crossing model (29): solid lines present the frequency modulation and the dashed lines indicate the corresponding parameter $\Delta_1$ ($= 4/3, 3, 5$ for curves 1, 2, 3, respectively).

The general Heun function involved in solution (9) presents a linear combination with constant coefficients of three Beta functions:

$$H_G = B_z(R,-1) + \frac{2\Delta_1(1-R)}{(\Delta_1+1)R} B_z(R+1,-1) + \frac{(\Delta_1-1)(R-1)}{(\Delta_1+1)(R+1)} B_z(R+2,-1), \quad (30)$$

where we have introduced the notation $R = \sqrt{4U_0^2 + \Delta_1^2} > 0$. Using the recurrence relation between consecutive neighbors

$$B_z(c,b) = \frac{z^c}{c}(1-z)^b + \frac{(b+c)}{c} B_z(c+1,b), \quad (31)$$

the sum in equation (30) is readily reduced to include just one Beta function. It is further checked that the coefficient of the term proportional to this beta function is zero so that the sum is finally simplified to a quasi-polynomial:

$$H_G = z^R \frac{(z-1)(1+R\Delta_1) - (z+1)(R+\Delta_1)}{R(R+1)(\Delta_1+1)(z-1)}, \quad (32)$$

The resultant solution of the two-state problem for $N = 2$ is eventually written as

$$a_2 = C_0 z^{\frac{\Delta_1+R}{2}} \left( (R-1)(\Delta_1-1) + \frac{2(R+\Delta_1)}{1-z} \right), \quad (33)$$

where $C_0$ is a constant that is defined from the initial conditions and

$$z(t) = \sqrt{\frac{\Delta_1+1}{\Delta_1-1}} e^{i(t-t_0)}. \quad (34)$$

We note that by changing $R$ to $-R$ we get the second independent solution of the problem.

Equation (33) shows that the *Floquet exponent* [27,28] for this solution is $i\lambda_2$ with $\lambda_2 = (\Delta_1 + R)/2 > 0$. This is immediately understood by noting that $z^{(\Delta_1+R)/2} = e^{i\lambda_2 t}$ while the term in brackets is $T = 2\pi/\Delta$ periodic like the coefficients of equation (2) are for the field configuration (29). By changing $R$ to $-R$ we get that the Floquet exponent for the second independent solution is $i\lambda_1$ with $\lambda_1 = (\Delta_1 - R)/2 < 0$. Finally, we note that $\lambda_{1,2}$ are the quasi-energies of the system for the constant field configuration $(U, \delta_t) = (U_0, \Delta_1)$. We conclude the discussion of the derived solution by examining the dynamics of the population of the second level described by equation (33). This dynamics is shown in figure 5. We note that since $1/(1-z)$ is the sum of the geometric progression with common ratio $z$:

$$\frac{1}{1-z} = \sum_{n=0}^{\infty} z^n, \quad (35)$$

and $R + \Delta_1 \neq 0$ for a non-zero interaction, it is understood that the time evolution of the excited level always involves all the harmonics of the driving field's frequency.

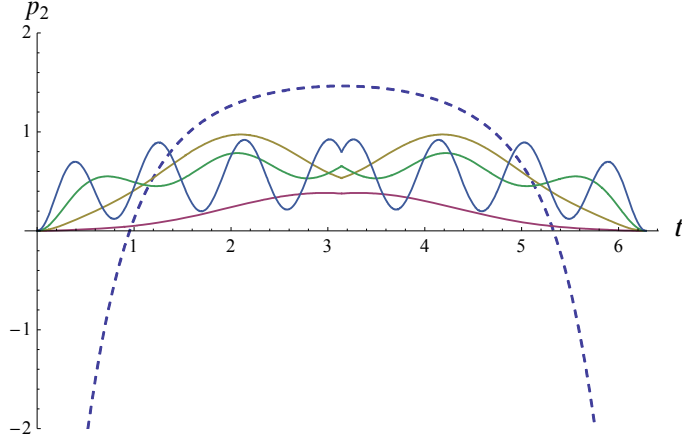

Fig. 5. The dynamics of the population of the excited level given by equation (33) for $U_0 = 0.3, 1, 2, 3.5$. Dashed line presents the detuning ($\Delta_1 = 2, \Delta = 1, t_0 = 0$).

## 6. Discussion

Thus, we have presented an analytic model of a quantum semi-classical time-dependent two-state problem belonging to the general Heun class. This is a constant-amplitude periodic level-crossing field configuration for which the general Heun function involved in the solution of the time-dependent Schrödinger equations can be expanded as a convergent series in terms of the incomplete Beta functions. The expansion is possible owing to the condition that a characteristic exponent of the general Heun equation for the regular singular point at infinity is zero. The coefficients of the expansion obey a three-term recurrence relation between the successive coefficients of the expansion that allows termination of the series.

Applying the termination, we have presented two constant amplitude periodic resonance crossing field configurations for which the solution is written in closed form as a finite sum of incomplete Beta functions. One of the models is an unconditionally integrable model and the other is a conditionally integrable one. For the unconditionally integrable model we have written down the explicit solution of the problem and have discussed the behavior of the system subject to excitation by corresponding field configuration. Notably, the solution in this case is finally simplified to an elementary function (quasi-polynomial). This function explicitly indicates the associated Floquet exponents, that is the spectrum of the quasi-energies.

**Acknowledgments**

This research has been conducted within the scope of the International Associated Laboratory IRMAS (CNRS-France & SCS-Armenia). The work has been supported by the Armenian State Committee of Science (SCS Grant No. 15T-1C323) and by the Armenian National Science and Education Fund (ANSEF Grant No. PS-4558). T.A. Ishkhanyan acknowledges the support from SPIE through a 2017 Optics and Photonics Education Scholarship and thanks the French Embassy in Armenia for a doctoral grant.

**References**

1. H. Nakamura, *Nonadiabatic Transition: Concepts, Basic Theories and Applications* (World Scientific, Singapore, 2012).
2. B.W. Shore, *The Theory of Coherent Atomic Excitation* (Wiley, New York, 1990).
3. L.D. Landau, "Zur theorie der energieubertragung. II", Phys. Z. Sowjetunion **2**, 46-51 (1932).
4. C. Zener, "Non-adiabatic crossing of energy levels", Proc. R. Soc. London, Ser. A **137**, 696-702 (1932).
5. E. Majorana, "Atomi Orientati in Campo magnetic Variable", Nuovo Cimento **9**, 43-50 (1932).
6. E.C.G. Stückelberg, "Theorie der unelastischen Stösse zwischen Atomen", Helv. Phys. Acta. **5**, 369-422 (1932).
7. E.E. Nikitin, "The probability of nonadiabatic transitions in the case of nondivergent terms", Opt. Spectrosc. **13**, 431-433 (1962).
8. Yu. N. Demkov and M. Kunike, "Hypergeometric model for two-state approximation in collision theory [in Russian]", Vestn. Leningr. Univ. Fis. Khim. **16**, 39 (1969).
9. F.T. Hioe, C.E. Carroll, "Analytic solutions to the two-state problem for chirped pulses", J. Opt. Soc. Am. B **3**, 497-502 (1985).
10. N. Rosen and C. Zener, "Double Stern-Gerlach experiment and related collision phenomena", Phys. Rev. **40**, 502-507 (1932).
11. A. Bambini and P.R. Berman, "Analytic solution to the two-state problem for a class of coupling potentials", Phys. Rev. A **23**, 2496-2501 (1981).
12. A.M. Ishkhanyan, "The integrability of the two-state problem in terms of confluent hypergeometric functions", J. Phys. A. **30**, 1203-1208 (1997).
13. A.M. Ishkhanyan, "New analytically integrable models of the two-state problem", Opt. Commun. **176**, 155-161 (2000).
14. A.M. Ishkhanyan, "New classes of analytic solutions of the three-state problem", J. Phys. A **33**, 5041-5547 (2000).
15. A. Ronveaux, *Heun's Differential Equations* (Oxford University Press, London, 1995).
16. S.Yu. Slavyanov and W. Lay, *Special functions* (Oxford University Press, Oxford, 2000).
17. F.W.J. Olver, D.W. Lozier, R.F. Boisvert, and C.W. Clark (eds.), *NIST Handbook of Mathematical Functions* (Cambridge University Press, New York, 2010).
18. A.M. Ishkhanyan, T.A. Shahverdyan, T.A. Ishkhanyan, "Thirty five classes of solutions of the quantum time-dependent two-state problem in terms of the general Heun functions", Eur. Phys. J. D **69**, 10 (2015).
19. A.M. Ishkhanyan and A.E. Grigoryan, "Fifteen classes of solutions of the quantum two-state problem in terms of the confluent Heun function", J. Phys. A **47**, 465205 (2014).

20. T.A. Shahverdyan, T.A. Ishkhanyan, A.E. Grigoryan, A.M. Ishkhanyan, "Analytic solutions of the quantum two-state problem in terms of the double, bi- and triconfluent Heun functions", J. Contemp. Physics (Armenian Ac. Sci.) **50**, 211-226 (2015).
21. A.M. Ishkhanyan and K.-A. Suominen, "Solutions of the two-level problem in terms of biconfluent Heun functions", J. Phys. A **34**, 6301-6306 (2001).
22. N. Svartholm, "Die Lösung der Fuchs'schen Differentialgleichung zweiter Ordnung durch Hypergeometrische Polynome", Math. Ann. **116**, 413 -421 (1939).
23. A. Erdélyi, "Certain expansions of solutions of the Heun equation", Q. J. Math. (Oxford) **15**, 62-69 (1944).
24. A.M. Manukyan, T.A. Ishkhanyan, M.V. Hakobyan, and A.M. Ishkhanyan, "A series solution of the general Heun equation in terms of incomplete Beta functions", IJDEA **13**(4), 231-239 (2014).
25. T.A. Ishkhanyan, T.A. Shahverdyan, A.M. Ishkhanyan, "Hypergeometric expansions of the solutions of the general Heun equation governed by two-term recurrence relations for expansion coefficients", arXiv:1403.7863 [math.CA] (2017).
26. A.M. Ishkhanyan, "The Appell hypergeometric expansions of the solutions of the general Heun equation", Constr. Approx. (2017) arXiv:1405.2871 [math-ph].
27. G. Floquet, "Sur les équations différentielles linéaires à coefficients périodiques", Annales scientifiques de l'École Normale Supérieure **12**, 47-88 (1883).
28. F. Bloch, "Über die Quantenmechanik der Elektronen in Kristallgittern," Zeitschrift für Physik A **52**, 555-600 (1929).